\documentclass[aps,prl,10pt,amsmath,floats,floatfix,twocolumn,superscriptaddress,altaffilletter,nofootinbib,shownopacs, frontmatterverbose,reprint]{revtex4-1}
\newcommand\prlsec[1]{\vspace{2mm} \textbf{\emph{#1}}\,---}
\usepackage{bm, graphics, graphicx, epsfig, soul, latexsym, hyperref ,multirow}
\usepackage{natbib}
\usepackage{comment}
\usepackage{hyperref}
\usepackage{subfigure}
\usepackage{graphicx,color}
\usepackage{dcolumn}
\usepackage{bm,amsmath, amssymb,}
\usepackage{mathrsfs}
\usepackage{bigints}
\usepackage{float}
\usepackage{orcidlink}
\usepackage{multirow}
\usepackage{xspace}
\usepackage{placeins}
\usepackage{subfigure}
\usepackage[normalem]{ulem}
\usepackage{booktabs}
\usepackage{tabularx}
\usepackage{esvect}
\usepackage{url}
\usepackage{makecell}

\usepackage{amsmath}

\newcommand{\be}{\begin{equation}}
\newcommand{\ee}{\end{equation}}

\newcommand{\eventname}{GW241011\xspace}
\newcommand{\fulltitle}{Testing the Kerr hypothesis beyond the quadrupole with \eventname}

\begin{document}

\title{Testing the Kerr hypothesis beyond the quadrupole with GW241011}
\author{Rimo Das\orcidlink{0009-0002-8388-0922}}
\email{rimo@physics.iitm.ac.in}
\affiliation{Department of Physics, Indian Institute of Technology Madras, Chennai 600036, India}
\affiliation{Centre for Strings, Gravitation and Cosmology, Department of Physics, Indian Institute of Technology Madras, Chennai 600036, India}
\author{N. V. Krishnendu\orcidlink{0000-0002-3483-7517}}
\email{k.naderivarium@bham.ac.uk}
\affiliation{School of Physics and Astronomy, University of Birmingham, Edgbaston, Birmingham, B15 2TT, UK}
\author{M. Saleem\orcidlink{0000-0002-3836-7751}} 
\email{muhammed.cholayil@austin.utexas.edu}\affiliation{Department of Physics, The University of Texas at Austin, 2515 Speedway, Austin, TX 78712, USA}
\author{Chandra Kant Mishra\orcidlink{0000-0002-8115-8728}}
\email{ckm@physics.iitm.ac.in}
\affiliation{Department of Physics, Indian Institute of Technology Madras, Chennai 600036, India}
\affiliation{Centre for Strings, Gravitation and Cosmology, Department of Physics, Indian Institute of Technology Madras, Chennai 600036, India}
\author{K. G. Arun\orcidlink{0000-0002-6960-8538}}
\email{kgarun@cmi.ac.in}
\affiliation{Chennai Mathematical Institute, Plot H1 SIPCOT IT Park, Siruseri, Tamilnadu 603103}
\affiliation{Max Planck Institute for Gravitational Physics (Albert Einstein Institute), D-30167 Hannover, Germany}
\affiliation{Leibniz Universit\"at Hannover, D-30167 Hannover, Germany}
\date{\today}

\begin{abstract}
All multipole moments of a Kerr black hole are uniquely determined by its mass and spin. Gravitational wave observations can test this prediction by measuring spin-induced multipole moments imprinted on the inspiral phase of compact binary mergers. In this \emph{Letter}, we show that the recently reported compact binary coalescence \eventname enables a simultaneous test of deviations in the spin-induced quadrupole and octupole moments of the binary components from their black hole values. We find no evidence for deviations from the Kerr prediction and place the first constraints on spin-induced octupole moments of the compact binary. This approach complements tests of the Kerr nature of compact binary merger remnants based on quasinormal mode measurements in the ringdown phase. 
\end{abstract}
\maketitle
\prlsec{Introduction}
\label{intro}
General relativity predicts that astrophysical black holes (BHs) are uniquely described by just their mass and spin~\cite{PhysRevD.40.3194,Carter71, Hansen74}, making them among the simplest macroscopic objects in the universe. Typically, they are produced by stellar collapses or compact object mergers. They can exist in pairs as demonstrated by the pathbreaking discovery of gravitational waves (GWs) from GW150914~\cite{LIGOScientific:2016aoc} whose properties are consistent with the gravitational radiation emitted by binary black hole (BBH) mergers. After that remarkable discovery, the LIGO-VIRGO-KAGRA (LVK) collaboration has detected $\mathcal{O}(200)$ confident binary merger events till the first part of the fourth observing run (O4a)~\cite{GWTC1-catalog,LIGOScientific:2021usb,KAGRA:2021vkt,GWTC4}. However, confidently establishing that the GW observed mergers are indeed BBH mergers is challenging~\cite{GWTC-1-TGR,GWTC-2-TGR,LIGOScientific:2021sio}. This is because there are various classes of exotic compact objects whose properties closely resemble those of BHs, at the level of precision currently achievable. This class of objects is collectively referred to as ``BH mimickers"~\cite{Giudice:2016zpa,Cardoso:2017cfl,Raposo:2018rjn,Maggio_2021,Cardoso:2019rvt}. 

 Quasi-normal modes emitted during the late ringdown of a compact binary coalescence have long been recognized as powerful probes of the black hole nature of the merger remnant~\cite{Dreyer:2003bv,BCW05}, as evidenced by recent results from LVK~\cite{TOGGW150914,GW250114_LVK_Area_Law}. However, the inspiral part of the gravitational waveform can provide complementary and unique probes of the black hole nature of the binary constituents (see, e.g.,~\cite{Cardoso:2017cfl,Datta:2019euh,Krishnendu:2017shb,Sennett:2017etc}). For binaries with spinning components, measurement of spin-induced multipole (SIM) moments is a robust method to test the Kerr nature~\cite{Krishnendu:2017shb, Saleem:2021vph,Krishnendu:2025rud} of the binary. This method makes use of the predictions of the ``no-hair" conjecture that all multipole moments of an astrophysical BH are completely described only by its mass and spin. The spin-induced quadrupole (SIQM) and spin-induced octupole moments (SIOM) of a BH are proportional to quadratic and cubic powers, respectively, of its dimensionless spin ${\boldsymbol \chi}=\frac{{\bf S}}{m^2}$ (in geometrical units). Therefore, when the inspiral dynamics of a compact binary is described using the post-Newtonian (PN) approximation--an expansion in the relative orbital velocity $v$~\cite{Blanchet:1995ez,Blanchet:1995fg,Blanchet:2002av}--these effects first appear at the orders where quadratic- and cubic-in-spin terms enter the phasing, namely at 2PN and 3.5PN orders~\cite{Marsat:2014xea,Bohe:2013cla,Marsat2015,Bohe:2015ana,Mishra:2016whh}, respectively.\footnote{$n$PN refers to terms proportional to $v^{2n}$ in the PN approximation.}.

By parametrizing these multipoles about their BH expressions, GW observations can be used to test the BBH nature of the compact binary systems~\cite{Krishnendu:2017shb, Krishnendu:2019tjp, Saini:2023gaw}. A method that exploits the measurement of the SIQM  parameter has been applied on the spinning BBHs  in all observing runs up to O4a and has yielded robust constraints on the BH nature of these systems~\cite{Krishnendu:2019tjp,Saleem:2021vph,GWTC-2-TGR,LIGOScientific:2026qni,LIGOScientific:2026fcf}. However, SIOM has not been measured from any of the detected events due to their low spins and shorter duration, which makes the measurement of this higher-order effect challenging~\cite{Saini:2023gaw}.

Recent detection of \eventname~\cite{LIGOScientific:2025brd} is novel due to its rapidly spinning primary (${\chi}_1=0.73^{+0.09}_{-0.09}$), the mass asymmetry (mass ratio $q=0.32^{+0.09}_{-0.08}$), and non-negligible spin precession (${\chi}_p=0.33^{+0.16}_{-0.13}$) in addition to the significant signal strength (network signal-to-noise (SNR) ratio of 37). \eventname provides the best constraint on the SIQM~\cite{LIGOScientific:2025brd} which have been used to constrain specific models of BH mimickers~\cite{Krishnendu:2025rud}. In this {\emph{Letter}}, we show how \eventname facilitates the first measurement of the effect of SIOM, which, in conjunction with the measurement of the SIQM, allows us to place the most stringent constraint to date on the BBH nature of the system.

\prlsec{Parametrizing departure from Kerr hypothesis}
For a non-BH compact object with mass $m$ and dimensionless spin $\boldsymbol{\chi}$, the SIQM and SIOM takes the form~\cite{Hansen74, Bohe:2013cla, Bohe:2015ana, Marsat2014},
\begin{eqnarray}
    Q=-\kappa\,\boldsymbol{\chi}^2\,m^3,\\ \label{eq;SIQM}
    O=-\lambda\,\boldsymbol{\chi}^3\, m^4,\label{eq:SIOM}
\end{eqnarray}
 where $\kappa$ and $\lambda$ are referred to as SIQM and SIOM parameters. For Kerr BHs,  $\kappa=\lambda=1$~\cite{Carter71,PhysRevLett.114.151102} according to the no-hair conjecture. For neutron stars and other BH mimickers, these parameters can differ from unity depending upon the internal composition~\cite{Ryan97, Vaglio:2022flq, Uchikata2015, Uchikata:2021jmy, Laar97, PappasMultipole2012}. In the inspiral part of the GW phasing, the SIQM terms appear at 2PN, 3PN, and 3.5PN orders, and the SIOM terms first appear at 3.5PN order. Therefore, we can parametrize the inspiral part of the BBH waveform, allowing for the possibility of $\kappa$ and $\lambda$ being different from unity, thereby facilitating a search for non-BBH signatures using GW data. 

The first Fisher matrix-based  forecast of the expected bounds on these parameters for Advanced LIGO and next-generation detectors was reported in \cite{Krishnendu:2017shb} and \cite{Krishnendu:2018nqa}, respectively. Later \cite{Krishnendu:2019tjp} applied the full Bayesian framework to perform the SIQM-based parameterized test using a parameterized singly precessing dominant mode  inspiral-merger-ringdown (IMR) waveform \texttt{IMRPhenomPv2}~\cite{Hannam:2013oca}. Ref.~\cite{Divyajyoti:2023izl} studied the improvement in the SIQM posteriors by parametrizing the doubly precessing IMR waveform \texttt{IMRPhenomXPHM}~\cite{Pratten:2020ceb}.  Specifically, \texttt{IMRPhenomXPHM} includes a few higher order spherical harmonics with $(\ell, |m|) = (2,1), (3,2), (3,3), (4,4)$ in addition to the dominant quadrupole mode $(\ell, |m|) = (2,2)$~\cite{Pratten:2020ceb}. The first study of the measurability of SIOM was reported in \cite{Krishnendu:2018nqa} in context of the next generation detectors using an inspiral model and also in \cite{Saini:2023gaw} using simulated loud injections, representing signals in the next-generation detectors, employing parametrized {\tt IMRPhenomPv2} waveforms. 

 Here we parameterize the \texttt{IMRPhenomXPHM} waveform model by implementing cubic-in-spin SIOM terms into the inspiral phase, besides the SIQM parameters, and perform a full Bayesian analysis. In this implementation, the deviations $\delta\kappa_i$ and $\delta\lambda_i$ are treated as free parameters, defined as $\kappa_i = 1 + \delta\kappa_i$ and $\lambda_i = 1 + \delta\lambda_i$, where $i=1,2$ label the binary constituents. In this way, we have cast the problem as a null test of BBH nature~\cite{Krishnendu:2019tjp,Saini:2023gaw}, allowing \emph{simultaneous} inference of SIOM and SIQM parameters using the state-of-the-art gravitational waveform \texttt{IMRPhenomXPHM}.  Having explained the method, we next apply it to \eventname.

\prlsec{Details of the analysis} 
Our goal is to estimate a set of parameters $\vec{\theta}$, which includes the standard BBH parameters as well as the SIM parameters, given the GW data $d$. In the Bayesian framework, we compute the multidimensional posterior distribution $p(\vec{\theta}\,|\,d,{\cal H})$, where ${\cal H}$ denotes the waveform model used for inference. Throughout this work, we use \texttt{IMRPhenomXPHM} for the standard BBH analyses and parametrized \texttt{IMRPhenomXPHM} for testing the BBH nature. According to Bayes' theorem, this posterior is proportional to the product of the likelihood function ${\cal L}(d\,|\,\vec{\theta},{\cal H})$ and the prior distribution $\pi(\vec{\theta})$.

We perform parameter estimation using the inference framework \texttt{bilby} and \texttt{bilby\_pipe}, with \texttt{dynesty} employed for sampling the likelihood~\cite{bilby_paper, dynesty, Romero-Shaw:2020owr}. For \texttt{IMRPhenomXPHM}, the parameter space comprises two component masses, six spin components, four parameters describing the sky location and orientation of the binary, the luminosity distance, and two kinematic parameters corresponding to the coalescence time and phase of the signal.  In the case of parametrized \texttt{IMRPhenomXPHM} inference, we infer SIM parameters in addition to the BBH ones, and obtain the posterior distributions of the SIM parameters by marginalizing over the standard BBH parameters.

Finally, for a given model, the Bayesian evidence ${\cal Z}$ is computed by integrating the product of the likelihood and prior over the full parameter space. If ${\cal Z}_\mathrm{BH}$ and ${\cal Z}_\mathrm{NBH}$ denote the Bayesian evidences for BBH ($\mathcal{H}_\mathrm{BH}$) and non-BBH ($\mathcal{H}_\mathrm{NBH}$) hypothesis, respectively, the Bayes factor comparing $\mathcal{H}_\mathrm{BH}$ to $\mathcal{H}_\mathrm{NBH}$ is defined as ${\cal B}^{\mathrm{BH}}_{\mathrm{NBH}} = {{\cal Z}_\mathrm{BH}}/{{\cal Z}_\mathrm{NBH}}$~\cite{Krishnendu:2019tjp,Krishnendu:2025rud}.  A sufficiently large value of ${\cal B}^{\mathrm{BH}}_{\mathrm{NBH}}$ indicates that the BBH hypothesis is strongly preferred over the non-BBH hypothesis by the data. 

\begin{figure}[t]
    \includegraphics[width= 3.5 in,height= 4.5 in]{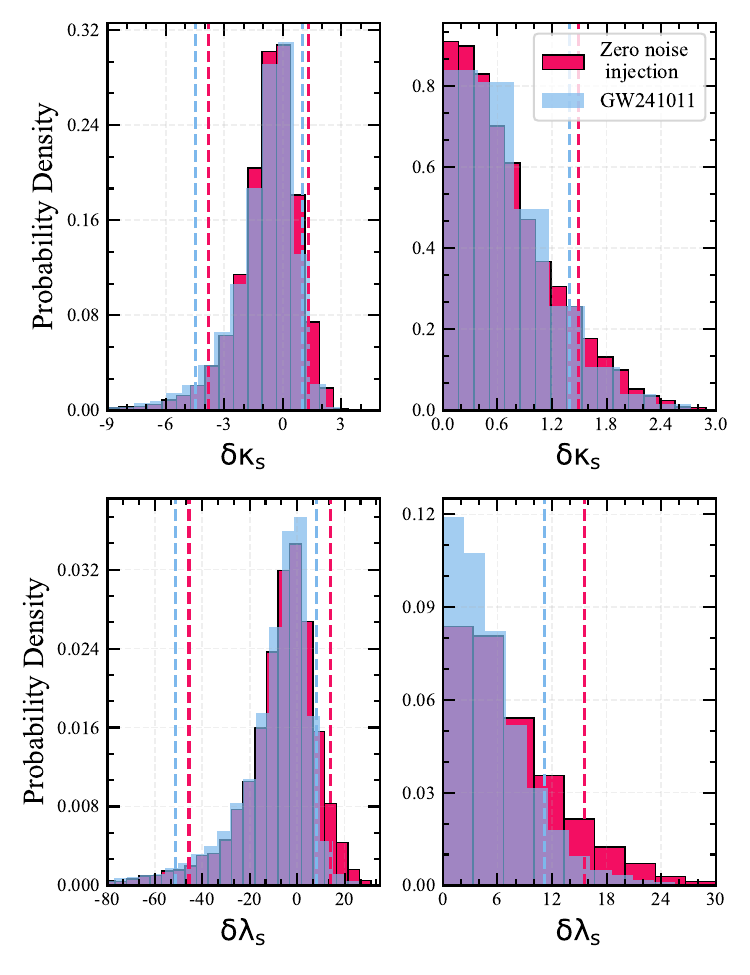}
    \caption{The 1-D marginalized distributions sky blue of $\delta\kappa_s$ and $\delta\lambda_s$  obtained from \eventname. The 1st (2nd) column shows the posteriors corresponding to the symmetric (positive) choice of priors. As expected, positive-only prior results are much tighter than the symmetric prior ones. The deep red histograms (in background) represent the $\delta\kappa_s$ and $\delta\lambda_s$ posteriors for \eventname-like zero noise injections. The sky blue dashed line represents the 90\% credible intervals (1st column) and 90\% upper bound (2nd column) for \eventname, respectively. Similarly, the deep red dashed line represents the 90\% bounds for zero noise injection. Further, 1-D marginalized posteriors of $\delta\lambda_s$ are wider than those of $\delta\kappa_s$, reflecting the fact that SIOM is a higher-order effect compared to SIQM. 
    }
    \label{fig:Ks_Ls}
\end{figure}
On the data of interest, either with a real astrophysical signal or a simulated signal, we perform a full Bayesian inference to estimate the posterior probability distributions of different combinations of $\delta\kappa_i$ and $\delta\lambda_i$, together with the standard BBH parameters ${\vec\theta}_{\rm BH}$.
In full generality, we have four free null parameters to be estimated alongside the BBH parameters: $\{\delta\kappa_1,\delta\kappa_2,\delta\lambda_1, \delta\lambda_2\}$. As $\delta\kappa$ and $\delta\lambda$ parameter pairs are highly degenerate among themselves as well as with other intrinsic BBH parameters such as spins and masses, a simultaneous estimation would return uninformative posteriors. Therefore, following \cite{Krishnendu:2017shb, Saini:2023gaw}, we infer the symmetric combinations of the SIQM and SIOM parameters, defined as $\delta\kappa_s=\frac{\delta\kappa_1+\delta\kappa_2}{2}$ and $\delta\lambda_s=\frac{\delta\lambda_1+\delta\lambda_2}{2}$ assuming the corresponding antisymmetric combinations $\delta\kappa_a$ and $\delta\lambda_a$ to be zero (implying $\delta\kappa_1=\delta\kappa_2$ and $\delta\lambda_1=\delta\lambda_2$). 

At the outset, this may look like a strong assumption as we are assuming the two compact objects share the same values of the SIQM and SIOM parameters. However, as we have cast this as a null test, we are looking for a statistically significant departure from the BBH nature. The posteriors of either (or both) of these parameters, excluding the BBH value of zero, should be considered as a potential signature of non-BBH nature, regardless of the assumption (see ~\cite{Krishnendu:2019tjp} for a detailed discussion). 
The Supplementary Material discusses two additional parameter choices and the results from them.

Next, we consider two scenarios for the priors on the SIM parameters. In the first, the priors on the $\delta\kappa$ and $\delta\lambda$ parameters are uniform and symmetric about zero (referred to as symmetric priors). In the second, they are uniform but restricted to strictly positive values (referred to as positive priors). The positive priors represent BH mimicker models such as boson stars~\cite{Ryan97,Ryan97b} for which these parameters are strictly positive, whereas the symmetric ones are more general and can also include objects such as gravastars~\cite{Uchikata:2021jmy,Uchikata2015,Uchikata2016} for which these parameters can be positive or negative.

 Further, we perform zero-noise injections of BBH signals using \texttt{IMRPhenomXPHM} with parameters and SNR similar to \eventname and analyse it with our framework for both the symmetric and positive priors on SIQM and SIOM parameters. A zero-noise injection refers to injecting a synthetic signal into data with no stochastic noise realization, while retaining the detector noise power spectral density in the likelihood. This allows us to assess the  parameter recoveries and degeneracies without being affected by statistical noise fluctuations. The data for \eventname is then analyzed for  the two prior choices. 

 At the end, a standard BBH analysis of the data using \texttt{IMRPhenomXPHM} and computing the corresponding evidence for Bayesian model selection between BBH and non-BBH hypotheses is performed.

\prlsec{Results}
Our main results are shown in Fig~\ref{fig:Ks_Ls}. It shows the one-dimensional marginalized posterior of $\delta \kappa_s$ and $\delta\lambda_s$, when they are simultaneously estimated, for the \eventname (sky blue) as well as the \eventname-like zero-noise injection (deep red). The left-hand (right-hand) column corresponds to the symmetric (positive) prior choices. The consistency between the zero-noise injection and the \eventname posteriors indicates that the inference is broadly consistent with expectations in this region of the parameter space. The posterior distributions of $\delta\kappa_s$ are significantly narrower than those of $\delta\lambda_s$, as expected since the SIOM is a higher-order effect. Nevertheless, the posteriors for $\delta\lambda_s$ remain highly informative owing to the large spin of the primary component of \eventname. For symmetric priors, the negative side of the prior range is less tightly constrained than the positive side for both parameters. This is a reflection of the degeneracy between SIM parameters and the component spin parameters. The correlation between these two sets of parameters is such that for binaries whose spins are aligned with the orbital angular momentum, the negative prior region is harder to constrain compared to the positive prior region. We refer the reader to Sec IIIB of ~\cite{Krishnendu:2019tjp} for a detailed discussion of this feature.

The posteriors from \eventname significantly constrain the family of non-BBH objects to those which have $-4.5\leq\delta\kappa_s\leq1.03$ ($\delta\kappa_s\leq 1.4$) and $-51.4\leq{ \delta\lambda_s}\leq 8.1$ ($\delta\lambda_s\leq11.2$) for symmetric (positive priors) at 90\% credibility (see Table~\ref{tab:BF}). These place the most stringent constraints on the non-Kerr nature of \eventname via simultaneous measurement of SIQM and SIOM parameters of the binary constituents. For the positive-prior case, we report only the 90\% upper bounds (see Table I in~\cite{Krishnendu:2019tjp}), as this corresponds to a one-sided test of deviations from the Kerr nature. 
 
Next, we perform Bayesian model selection between the BBH and non-BBH hypotheses by computing the logarithmic Bayes factors (base 10) for all cases considered. These are reported in Table~\ref{tab:BF}. The log Bayes factors obtained from the simultaneous measurement of $\delta\kappa_s$ and $\delta\lambda_s$ is of the order of 70 for different prior choices. Such large Bayes factors indicate that the data strongly favor the BBH hypothesis over the non-BBH hypothesis for \eventname. Given the generality of the parameterization---which incorporates two leading SIM moments of the compact objects---and the magnitude of the Bayes factors, these results provide the most stringent evidence to-date for the BBH nature of a GW source. They may also be interpreted as the first observational test of the Kerr hypothesis extending beyond the quadrupole order using the inspiral part of the signal.

There are several alternative parameterizations that can be, albeit at the cost of stronger assumptions. One such approach involves the simultaneous estimation of $\delta\kappa_1$ and $\delta\lambda_1$, assuming that the secondary object is a BH rather than that the primary and secondary share the same values of these parameters. We find this yields  significantly less informative posteriors on this event compared to those from the $\{\delta\kappa_s,\delta\lambda_s\}$ parameterization (see entries in Table~\ref{tab:BF} and the corresponding posteriors in the Supplementary Material). Another possible parameterization involves estimating only $\delta\lambda_s$ while fixing $\delta\kappa_s = 0$, as discussed in ~\cite{Saini:2023gaw}. In this case, the test probes the consistency of the spin-induced octupole moment under the assumption that the object satisfies the Kerr hypothesis at the quadrupolar order. The detailed results of these tests are also presented in the Supplementary Material. The corresponding posterior widths and Bayes factors are summarized in Table~\ref{tab:BF}. All of these analyses support the conclusion drawn above from Fig.~\ref{fig:Ks_Ls}, namely that the \eventname data strongly favor the BBH hypothesis.
\begin{table}[t]
    \centering
    \renewcommand{\arraystretch}{2.2}
    \begin{tabular}{c c c c}
    \hline 
       \textbf{\makecell{Parameters}}  & \textbf{\makecell{Prior range}} & \textbf{\makecell{Inferred Value}}& \textbf{\makecell{Bayes factor \\ ($\rm{log}_{10}\mathcal{B}^{\rm BH}_{\rm NBH}$)}}\\
      \hline
        \{$\delta\kappa_s, \delta\lambda_s$\} & $\mathcal{U}[-500, 500]$ & \makecell{$\delta\kappa_s=-0.6_{-3.8}^{+1.6}$,\\        $\delta\lambda_s=-5.8_{-45.6}^{+13.9}$} & 71.1\\
      \{$\delta\kappa_s, \delta\lambda_s$\} & $\mathcal{U}[0, 500]$ & \makecell{$\delta\kappa_s~\leq1.4$,\\ $\delta\lambda_s~\leq11.2$}& 71.2\\  
     \{$\delta\kappa_1, \delta\lambda_1$\} & $\mathcal{U}[-500, 500]$ & \makecell{$\delta\kappa_1=-9.1_{-233.01}^{+12.1}$,\\        $\delta\lambda_1=-33.8_{-409.6}^{+403.3}$}& 76.1\\    
       \{$\delta\kappa_1, \delta\lambda_1$\} & $\mathcal{U}[0, 500]$ & \makecell{$\delta\kappa_1~\leq3.8$,\\ $\delta\lambda_1~\leq99.4$}& 80.7\\
       \{$\delta\lambda_s$\} & $\mathcal{U}[-500, 500]$ & \makecell{$\delta\lambda_s=-0.5_{-0.8}^{+1.04}$}&64.4\\
       \{$\delta\lambda_s$\} & $\mathcal{U}[0, 500]$ & \makecell{$\delta\lambda_s~\leq1.3$}&65.9\\
       \{$\delta\lambda_1$\} & $\mathcal{U}[-500, 500]$ & \makecell{$\delta\lambda_1=-0.5_{-0.9}^{+1.8}$}&67.5\\
       \{$\delta\lambda_1$\} & $\mathcal{U}[0, 500]$ & \makecell{$\delta\lambda_1~\leq2.01$}&67.6\\
       \hline
    \end{tabular}
    \caption{Details of priors and inferred values (median and $90\%$ credible intervals) and Bayes factors for different combinations of SIOM and SIOM parameters. For positive only prior, cases we quote the 90\% upper bounds for the inferred parameter. We have choosen uniform prior for all the above cases.}
    \label{tab:BF}
\end{table}

\prlsec{Conclusion and Discussion} 
It has long been argued, primarily in the context of LISA, that GW observations of compact binaries can probe the multipolar structure of the constituent compact objects~\cite{Ryan97}. Here, we investigate an analogous scenario for ground-based detectors, where the highly spinning compact binary \eventname enables the simultaneous measurement of the spin-induced quadrupole and octupole moments, thereby providing strong evidence for the BBH nature of the system with high statistical significance. This measurement enables, for the first time, a test of the Kerr hypothesis beyond the quadrupole order  using the inspiral part of the gravitational waveform, and places stringent constraints on deviations at the octupolar order.

Our analysis employs a suite of complementary parameterizations, all of which consistently confirm the BBH hypothesis for \eventname, extending and complementing earlier tests based solely on the SIQM~\cite{GW241011,Krishnendu:2025rud}.  These tests are complementary to ringdown-based probes of the Kerr nature of the merger remnant~\cite{Dreyer:2003bv,BCW05}, recently applied to GW250114~\cite{LIGOScientific:2025wao}. While ringdown measurements constrain the properties of the remnant object, inspiral-based tests such as ours provide direct constraints on the nature of the progenitor compact objects.

As GW detector sensitivities continue to improve, the growing population of \eventname-like systems will enable increasingly stringent tests of the BBH nature of compact binaries~\cite{Saleem:2021vph}. In particular, future high-SNR events may permit the measurement of higher-order spin effects, such as the spin-induced hexadecapole moment arising from quartic-in-spin contributions to the GW phasing~\cite{Levi:2015msa,Levi:2020lfn,Khalil:2023kep}. Given that the fourth observing run already includes events that are two to three times louder than those from earlier runs~\cite{GW250114_LVK_Area_Law,LIGO,LIGOScientific:2025cmm}, the prospects for detecting highly spinning systems with SNRs of the order $\sim100$ or greater in forthcoming observing runs are promising. Such events are expected to be common in the data of next-generation ground-based observatories, including Cosmic Explorer~\cite{CE} and Einstein Telescope~\cite{ET:Sathya}. These observations could potentially enable precision tests of the Kerr hypothesis beyond the octupole order or reveal signatures of BH mimickers. We therefore end by emphasizing that our results strongly motivate continued theoretical efforts toward computing higher-order post-Newtonian contributions (4PN and beyond) for spinning compact binaries.

\prlsec{Acknowledgments}
We thank Ish Gupta for his review and feedback on the manuscript. We thank Ajit Mehta, Prashant Kocherlakota and Nathan Johnson-McDaniel for their useful comments on the draft. Rimo Das and Chandra Kant Mishra acknowledge Core Research Grant No. CRG/2022/007959. 
Krishnendu is supported by STFC grant ST/Y00423X/1. MS acknowledge the support from Weinberg Institute for Theoretical Physics at University of Texas at Austin.  K.G.A. is supported by Advance Research Grant ANRF/ARG/2025/000931/PS of Anusandhan National Research Foundation and Max Planck Society. K. G. Arun also acknowledges a grant from Infosys Foundation. 

\begin{figure}[t]
    \includegraphics[width= 3.3 in]{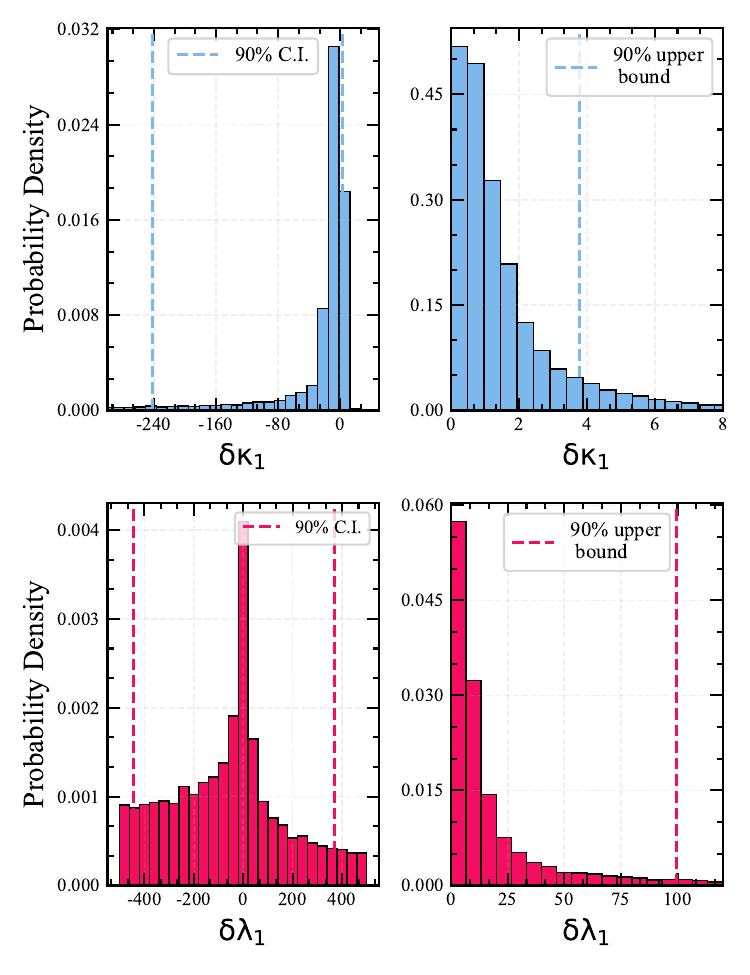}
    \caption{The first and second column shows 1-D marginalized distributions of $\delta\kappa_1$ (sky blue) and $\delta\lambda_1$ (deep red) for \eventname assuming symmetric and positive only prior. Here, both the parameters ($\delta\kappa_1$ and $\delta\lambda_1$) are measured simultaneously with the BBH parameters. The ranges for symmetric and positive only priors are taken to be uniform in [-500,500] and [0,500], respectively. }
    \label{fig:dK1_dL1}
\end{figure}

\begin{figure}[t]
    \includegraphics[width= 3.3 in]{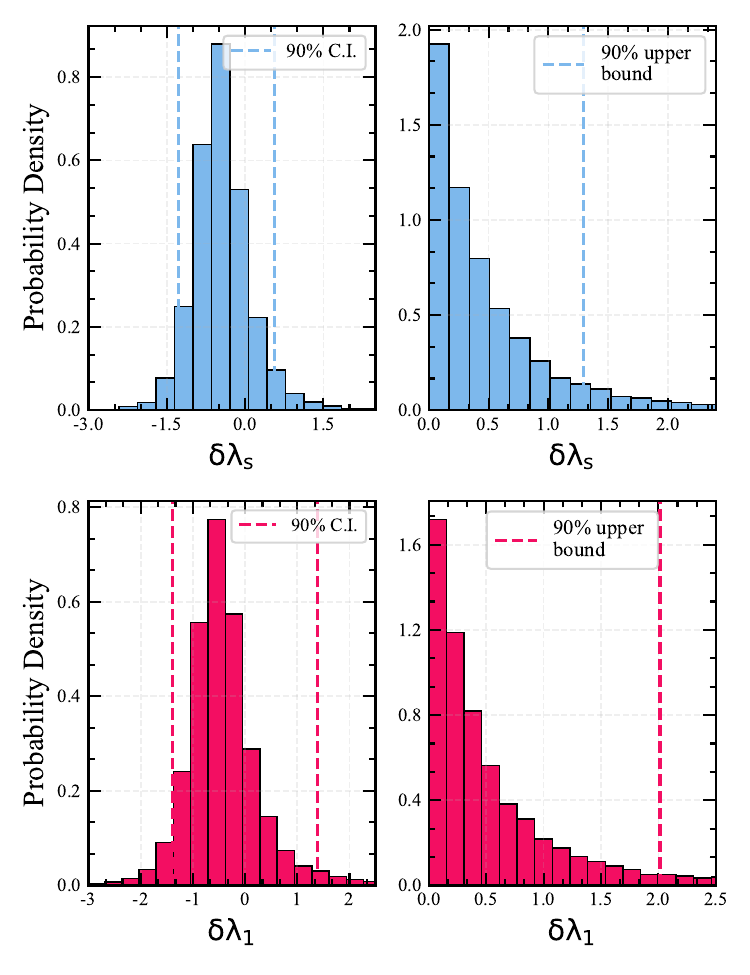}
    \caption{The posteriors shows single-parameter measurement of $\delta\lambda_s$ and $\delta\lambda_1$ with BBH parameters. The first and second column shows 1-D marginalized distributions of $\delta\lambda_s$ (sky blue) and $\delta\lambda_1$ (deep red) for \eventname assuming symmetric and positive only prior. The ranges for symmetric and positive only priors are taken to be uniform in [-500,500] and [0,500], respectively.}
    \label{fig:dLs_dL1}
\end{figure}

The authors are grateful
for computational resources provided by the LIGO Laboratory and supported by National Science Foundation Grants PHY0757058 and PHY-0823459 and the Symmetry cluster at Perimeter Institute. This research has made use of data or software obtained from the Gravitational Wave Open Science Center~\cite{GWOSC}, a service of the LIGO
Scientific Collaboration, the Virgo Collaboration, and KAGRA. This material is based upon work supported by NSF's LIGO Laboratory which is a major facility fully funded by the National Science Foundation. This document has LIGO preprint number {\tt LIGO-P2600147}.
\bibliography{ref}
\clearpage
{\centering
\begin{center}{\large Supplementary Material\\[3ex] \textbf{\fulltitle}}\end{center}
}
In this section, we discuss two alternative parametrizations we explored in the context of \eventname, results of which can be found in Table~\ref{tab:BF} as well as in Fig.~\ref{fig:dK1_dL1} and Fig.~\ref{fig:dLs_dL1}. 

\prlsec{Constraints on  \boldsymbol{$\delta\kappa_1$} and \boldsymbol{$\delta\lambda_1$} from simultaneous measurements}
As discussed in the main section, several parameterizations are possible depending on different assumptions on the nature of the compact objects. Fig.~\ref{fig:dK1_dL1} shows the posteriors of SIQM and SIOM parameters for \eventname measured at the same time, assuming the secondary compact object is a BH (i.e. $\delta\kappa_2= \delta\lambda_2=0$). Under this assumption we can bound the parameters in range  $-242\leq\delta\kappa_1\leq3.1~(\delta\kappa_1\leq3.8)$ and $-443\leq\delta\lambda_1\leq369~(\delta\lambda_1\leq99.4)$ with 90\% credibility for symmetric (positive) prior cases.

 Comparing these results with those in Fig.~\ref{fig:Ks_Ls}, it is evident that the posteriors on $\delta\kappa_s$ and $\delta\lambda_s$ are significantly tighter than those obtained from the simultaneous measurement of $\delta\kappa_1$ and $\delta\lambda_1$, indicating that the latter parametrization provides less constraining power on the SIQM and SIOM parameter. This difference in measurement uncertainty arises from the structure of the 3.5PN phase term, where SIQM and SIOM contributions enter together with the mass and spin parameters. When sampling over $\delta\kappa_1$ and $\delta\lambda_1$ under the assumption that the secondary is a BH, the phasing terms involving $\delta\kappa_2$ and $\delta\lambda_2$ are set to zero. In contrast, sampling over $\delta\kappa_s$ and $\delta\lambda_s$ assumes $\delta\kappa_a = 0$ and $\delta\lambda_a = 0$. These two choices lead to different phase contributions, which in turn modify how the SIQM and SIOM parameters correlate with each other and with the binary’s mass and spin parameters. 
 
  More specifically, starting from the parametrized phasing expressions in the Supplementary Material of \cite{Krishnendu:2017shb}, one finds that the phasing for the $\{\delta\kappa_2 \to 0, \delta\lambda_2 \to 0\}$ parametrization immediately kills all self-spin terms that are quadratic or cubic in spins, as they always accompany $\delta\lambda_2$ and $\delta\kappa_2$ terms, respectively. This is unlike the $\{\delta\kappa_a\rightarrow 0,\delta\lambda_a\rightarrow 0\}$ parametrization in which $\delta\kappa_s$ and $\delta\lambda_s$ terms partially retain these terms. This significantly limits the ability of the $\{\delta\kappa_1,\delta\lambda_1\}$ parametrization to yield constraints comparable to those obtained with the $\{\delta\kappa_s,\delta\lambda_s\}$ parametrization. A similar trend was noted in \cite{GW241011} in the context of single-parameter tests, where posteriors from $\delta\kappa_s$ were more informative than those of $\delta\kappa_1$, which can also be explained by this argument. Despite the broader posteriors, the Bayes factors of both parameterizations are comparable, offering a stringent test of the Kerr nature of the primary compact object.

\prlsec{Single-parameter tests of \boldsymbol{$\delta\lambda_1$} and \boldsymbol{$\delta\lambda_s$}}
 In this scenario, the null test probes only the octupolar parameter under the assumption that the Kerr hypothesis is already satisfied at the quadrupolar level. First, we constrain $\delta\lambda_s$ assuming both the component objects of the same type (i.e. $\delta\lambda_1=\delta\lambda_2$). This is presented in the top row of Fig.~\ref{fig:dLs_dL1} for symmetric and positive prior choices. In the second case, we constrain $\delta\lambda_1$ by fixing $\delta\lambda_2$ to the BH value as shown in the bottom panel of Fig.~\ref{fig:dLs_dL1}, again for the two prior choices.
  
As we can observe in Fig.~\ref{fig:dLs_dL1}, the posteriors for $\delta\lambda$ are more stringent for the one-parameter case compared to the two-parameter measurement, as one would expect. We can constrain the $\delta\lambda$ parameters in range $-1.4\leq\delta\lambda_1\leq1.4~(\delta\lambda_1\leq2.01)$ and $-1.3\leq\delta\lambda_s\leq0.6~(\delta\lambda_1\leq1.3)$ with 90\% credibility for symmetric (positive) prior cases. The $\delta\lambda_s$ parameterization results in tighter bounds ($\sim1.5$ times) compared to the $\delta\lambda_1$ parameterization, consistent with our previous findings where symmetrized parameters yielded more informative posteriors in Fig~\ref{fig:Ks_Ls}. Moreover, the values of the log Bayes factor for single parameter measurement are of the order of 60,  expected due to the lower dimensionality of the parameter space compared to the two parameter case. Such large values of log Bayes factors in conjunction with the stringent bound on $\delta\lambda$ parameter establish the strong preference for the BBH nature of \eventname even beyond the quadrupole order,  regardless of the parametrization employed.

\end{document}